\documentclass{article}
\usepackage{amsmath}
\usepackage{amsfonts}
\usepackage{amssymb}
\usepackage{hyperref}

\begin{document}

\title{\textbf{Two }$\mathbf{\theta }_{\mu \nu }$\textbf{-deformed covariant
relativistic quantum phase spaces as Poincare-Hopf algebroids}}
\author{\textbf{Jerzy Lukierski}, \textbf{Mariusz Woronowicz} \vspace{12pt} 
\\
Institute for Theoretical Physics, University of Wroclaw,\\
pl. Maxa Borna 9, 50-205 Wroclaw, Poland }
\date{}
\maketitle

\begin{abstract}
We consider two quantum phase spaces which can be described by two Hopf
algebroids linked with the well-known $\theta _{\mu \nu }$-deformed $D=4$
Poincare-Hopf algebra $\mathbb{H}$. The first algebroid describes $\theta
_{\mu \nu }$-deformed relativistic phase space with canonical NC space-time
(constant $\theta _{\mu \nu }$ parameters) and the second one incorporates
dual to $\mathbb{H}$ quantum $\theta _{\mu \nu }$-deformed Poincare-Hopf
group algebra $\mathbb{G}$, which contains noncommutative space-time
translations given by $\Lambda $-dependent $\Theta _{\mu \nu }$ parameters ($%
\Lambda $ $\equiv \Lambda _{\mu \nu }$ parametrize classical Lorentz group).
The canonical $\theta _{\mu \nu }$-deformed space-time algebra and its
quantum phase space extension is covariant under the quantum Poincare
transformations described by $\mathbb{G}$. We will also comment on the use
of Hopf algebroids for the description of multiparticle structures in
quantum phase spaces.
\end{abstract}

\section{Introduction}

There have been proposed in recent years various models of noncommutative
(NC) space-times which characterizes space-time geometry if the quantum
gravity (QG) effects are included (see e.g. \cite{e1}-\cite{e8}). In this
paper we shall study the canonical case of quantum space-times, with NC
counterparts $\hat{x}_{\mu }~(\mu =0,1,2,3,4)$ of space-time coordinates
satisfying the well-known formula 
\begin{equation}
\lbrack \hat{x}_{\mu },\hat{x}_{\nu }]=i\lambda ^{2}\theta _{\mu \nu },
\label{kom}
\end{equation}%
where $\theta _{\mu \nu }=-\theta _{\nu \mu }$ is a numerical $4\times 4$
matrix\footnote{%
In this paper we shall consider the case $D=3+1$} and $[\lambda ]=[l]$
describes an elementary length, which for QG-generated noncommutativity is
linked with Planck length 
\begin{equation}
\lambda _{P}=\frac{\hbar }{m_{P}c}=\sqrt{\frac{\hbar G}{c^{3}}},  \label{len}
\end{equation}%
where $m_{P}$ is the Planck mass, and $G$ describes the Newton constant
characterizing the gravitational interactions. If $\lambda $ is proportional
to $\lambda _{P}$, from formula (\ref{len}) due to the presence of Planck
constant $\hbar $, one can deduce the quantum-mechanical and QG origin of
relation (\ref{kom}) $(\lambda _{P}=0$ if $\hbar \rightarrow 0$ or $%
G\rightarrow 0)$. Further for simplicity we shall use the choice of units $%
\hbar =c=1$.

The well-known $\theta _{\mu \nu }$-deformed quantum space-times (see (\ref%
{kom})) and associated quantum phase spaces are generated by the following
Abelian twist \cite{oeck},\cite{chaichian}%
\begin{equation}
\mathcal{F}\equiv \mathcal{F}_{(1)}\otimes \mathcal{F}_{(2)}=\exp \left( 
\frac{i}{2}\lambda ^{2}\theta ^{\mu \nu }p_{\mu }\otimes p_{\nu }\right) .
\label{twistF}
\end{equation}%
It defines $\theta _{\mu \nu }$-deformed quantum Poincare-Hopf algebra $%
\mathbb{H}$, with nondeformed Poincare algebra sectors however with modified
coproducts and antipodes \cite{drin} 
\begin{eqnarray}
\Delta _{\mathcal{F}}(h) &=&\mathcal{F}\circ \,\Delta _{0}(h)\circ \mathcal{F%
}^{-1},\qquad h=\{p_{\mu },M_{\mu \nu }\},\,  \label{ko1} \\
S_{\mathcal{F}}(h) &=&U\,S_{0}(h)\,U^{-1},\qquad U=\mathcal{F}_{(1)}S_{0}(%
\mathcal{F}_{(2)}).  \label{ko2}
\end{eqnarray}

The twist (\ref{twistF}) can be employed in two ways:

\begin{enumerate}
\item[i)] We introduce the classical Minkowski space-time coordinates $%
x_{\mu }\in \mathbf{X}$ as vectorial representation of relativistic
symmetries described by classical Poincare algebra $\mathcal{P}$, with the
algebraic structure of $\mathcal{P}\otimes \mathbf{X}$ given by the
semidirect product $\mathcal{P}\rtimes \mathbf{X}$. In quantum-deformed
theory one introduces the NC space-times $(\hat{x}_{\mu }\in $ $\mathbf{\hat{%
X}}\mathbb{)}$ as the module algebra (NC algebraic representation) of
quantum Poincare-Hopf algebra $\mathbb{H}$. In the case of $\theta _{\mu \nu
}$-deformation the quantum space-time coordinates $\hat{x}_{\mu }\in \mathbf{%
\hat{X}}$ are obtained from Drinfeld twisting procedure \cite{drin} by star
product technique \cite{bloh},\cite{abp}.

Let us consider the algebra $\hat{A}$ of functions on $\mathbf{\hat{X}}$ $(f(%
\hat{x}\mathbb{)\in }\hat{A})$. In the scheme of twist quantization one can
represent the algebra $(\hat{A},\cdot )$ by the algebra $(A,\star _{\mathcal{%
F}})$ of classical functions, with their multiplication defined by the
nonlocal star product $\star _{\mathcal{F}}$\footnote{%
Because the twist $\mathcal{F}$ is the function of classical Poincare
algebra generators $\hat{g}=(p_{\mu },M_{\mu \nu })$, the action $\hat{g}%
\vartriangleright f(x)$ in formula (\ref{four}) is described by the
differential realization of classical Poincare algebra on functions of
standard Minkowski coordinates $x_{\mu }$.} 
\begin{equation}
f(\hat{x})\cdot g(\hat{x})\simeq f(x)\star _{\mathcal{F}}g(x)=m[\mathcal{F}%
^{-1}\circ (f\otimes g)]=(\mathcal{F}_{(1)}^{-1}\rhd f)(\mathcal{F}%
_{(2)}^{-1}\rhd g).  \label{four}
\end{equation}

Putting in (\ref{four}) $f(\hat{x})=\hat{x}_{\mu },g(\hat{x})\equiv 1$ one
gets 
\begin{equation}
\hat{x}_{\mu }=m[\mathcal{F}^{-1}(\vartriangleright \otimes 1)(x_{\mu
}\otimes 1)]=(\mathcal{F}_{(1)}^{-1}\rhd x_{\mu })\mathcal{F}_{(2)}^{-1}.
\label{ncco}
\end{equation}%
Formula (\ref{ncco}) provides the quantum map expressing the deformed NC
space-time coordinates by a nonlocal map in undeformed relativistic phase
space $(x_{\mu },p_{\mu }=\frac{1}{i}\partial _{\mu })$. If we put $%
f(x)=x_{\mu },g(x)\equiv x_{\nu }$ one can calculate from (\ref{four}) the
commutator given by the formula (\ref{kom}).%
\begin{equation}
\left[ \hat{x}_{\mu },\hat{x}_{\nu }\right] \simeq \left[ x_{\mu },x_{\nu }%
\right] _{\star _{\mathcal{F}}}\equiv x_{\mu },\star _{\mathcal{F}}x_{\nu
}-x_{\nu }\star _{\mathcal{F}}x_{\mu }=i\lambda ^{2}\theta _{\mu \nu }.
\label{st}
\end{equation}

\item[ii)] By Hopf-algebraic duality one can define the $\theta _{\mu \nu }$%
-deformed quantum Poincare group $\mathbb{G}$, with generators $\hat{g}=\{%
\hat{\xi}_{\mu },\hat{\Lambda}_{\mu \nu }\}$, describing generalized NC
coordinates on algebraic $\theta _{\mu \nu }$-deformed Poincare group
manifold \cite{koss} with quantum Lorentz group parameters $\hat{\Lambda}%
_{\mu \nu }$ and the NC quantum Poincare group translations $\hat{\xi}_{\mu
} $ satisfying the following algebra%
\begin{equation}
\left[ \hat{\xi}_{\mu },\hat{\xi}_{\nu }\right] =i\lambda ^{2}\theta ^{\rho
\sigma }(\eta _{\mu \rho }\eta _{\nu \sigma }-\hat{\Lambda}_{\mu \rho }\hat{%
\Lambda}_{\nu \sigma })\,:=i\lambda ^{2}\Theta _{\mu \nu }(\hat{\Lambda}).
\label{lnc}
\end{equation}

Using the Heisenberg double construction \cite{majid},\cite{nowluk} given by
particular choice of semidirect product $\mathbb{G\rtimes H}$ called smash
product, one obtains $(10+10)$-dimensional generalized $\theta _{\mu \nu }$%
-deformed quantum phase space $\mathcal{H}^{(10+10)}=(\hat{\xi}_{\mu },\hat{%
\Lambda}_{\mu \nu },p_{\mu },M_{\mu \nu })$. Such phase space can be
employed in physical applications for the description of NC dynamics on
algebraic $\theta _{\mu \nu }$-deformed quantum Poincare group manifold%
\footnote{%
The idea of phase space description of the dynamics on the classical group
and coset group manifolds is due to Souriau \cite{sou} and Kostant \cite{kos}%
.}.
\end{enumerate}

It appears that both $\theta _{\mu \nu }$-deformed structures presented
above (see (\ref{kom}) and (\ref{lnc})) are necessary in order to describe
in complete way the NC space-time (\ref{st}) as describing quantum
Poincare-covariant $\mathbb{H-}$module, which transform under quantum
Poincare group $\mathbb{G}$ in the following standard way%
\begin{equation}
\hat{x}_{\mu }^{\prime }=\hat{\Lambda}_{\mu \nu }\hat{x}^{\nu }+\hat{\xi}%
_{\mu },  \label{poin}
\end{equation}%
where NC translations $\hat{\xi}_{\mu }\in \mathbf{T}$ satisfy the relation (%
\ref{lnc}). We shall show that to describe the quantum Poincare
transformations (\ref{poin}) using star-product formula one should extend
the star multiplication (\ref{four}) to the functions of NC variables $\hat{x%
}_{\mu }$ (NC Minkowski space) and $\hat{\xi}_{\mu }$ (NC Poincare
translations) (see \cite{koss})\footnote{%
See \cite{klm} for the use of star product to represent the quantum group
transformations.}.

The plan of our paper is following. In Sect. 2 we recall the $\theta _{\mu
\nu }$-deformed Poincare-Hopf algebra $\mathbb{H}$ extended by NC space-time
coordinates $\hat{x}_{\mu }\in \mathbf{\hat{X}}$ which describe the Lorentz
group extension of the relativistic phase space $(\hat{x}_{\mu },p_{\mu
})\in \mathbf{\hat{X}}\mathbb{\rtimes }\mathcal{T}^{4}$%
\begin{equation}
\mathbf{\hat{X}}\mathbb{\rtimes H=\mathbf{\hat{X}}\rtimes }(\mathcal{T}^{4}%
\mathbb{\rtimes }O(3,1)\mathbb{)}.
\end{equation}%
In Sect. 3 we describe the $\theta _{\mu \nu }$-deformed Poincare quantum
group $\mathbb{G}$ and calculate the $\theta _{\mu \nu }$-deformed
Heisenberg double $\mathcal{H}^{(10+10)}=\mathbb{H}\rtimes \mathbb{G}$, with
generalized NC\ Poincare coordinates $\mathbb{\{}\hat{\xi}_{\mu },\hat{%
\Lambda}_{\mu \nu }\}\in \mathbb{G}$ and generalized momenta $\{p_{\mu
},M_{\mu \nu }\}\in \mathbb{H}$. In Sect. 4 we shall derive the covariance
under the quantum Poincare group transformations given by formula (\ref{poin}%
). In Sect. 5 we specify the data which define two Hopf algebroids, first
providing the quantum Poincare-covariant $\theta _{\mu \nu }$-deformed phase
space $(\hat{x}_{\mu },p_{\mu })\in \mathbf{\hat{X}}\mathbb{\rtimes }%
\mathcal{T}^{4}$ with positions (coordinates) described by algebra $\mathbf{%
\hat{X}}$ $\mathbb{(}\hat{x}_{\mu }\mathbb{\in \mathbf{\hat{X}})}$
supplemented with Lorentz transformations (Lorentz parameters $\Lambda _{\mu
\nu }$) and the second Hopf algebroid describing quantum $\theta _{\mu \nu }$%
-deformed Poincare symmetry transformations (see (\ref{poin})), with
coordinate sector described by quantum $\theta _{\mu \nu }$-deformed
Poincare group $\mathbb{G}$. In particular in Sect. 5.4. by following
earlier applications of Hopf algebroids to the description of $\kappa $%
-deformed quantum phase spaces \cite{lws1},\cite{lws} we shall consider the
applicability of coalgebra sector of $\theta _{\mu \nu }$-deformed Hopf
algebroids to the phase space description of the multi-particle states.

\section{ $\protect\theta _{\protect\mu \protect\nu }$-deformed quantum
Poincare algebra $\mathbb{H}$ and NC space-time as $\mathbb{H}$-module}

\subsection{\protect\bigskip Twist deformed quantum Poincare algebra $%
\mathbb{H}$}

The classical $D=4$ Poincare-Hopf algebra looks as follows%
\begin{align}
\lbrack p_{\mu },p_{\nu }]& =0  \notag \\
\lbrack M_{\mu \nu },p_{\rho }]& =i(\eta _{\nu \rho }p_{\mu }-\eta _{\mu
\rho }p_{\nu })  \label{poinc} \\
\lbrack M_{\mu \nu },M_{\rho \sigma }]& =i(\eta _{\nu \rho }M_{\mu \sigma
}-\eta _{\mu \rho }M_{\nu \sigma }-\eta _{\nu \sigma }M_{\mu \rho }+\eta
_{\mu \sigma }M_{\nu \rho })  \notag
\end{align}%
where $\eta _{\mu \nu }=diag(-1,1,...,1)$ and is supplemented by primitive
costructure maps 
\begin{equation}
\Delta _{0}(h)=h\otimes 1+1\otimes h,\qquad S_{0}(h)=-h,\qquad \epsilon
_{0}(h)=0.  \label{undef-antip}
\end{equation}

The twist $\mathcal{F}$ is an element of\thinspace\ $\mathbb{H}\otimes 
\mathbb{H}$ $(\mathbb{H=}\mathcal{U(P)})$ which has an inverse, satisfies
the cocycle condition $\mathcal{F}_{12}\,\left( \Delta _{0}\otimes 1\right)
\,\mathcal{F}=\mathcal{F}_{23}\,\left( 1\otimes \Delta _{0}\right) \,%
\mathcal{F}$ $\,$and the normalization condition $(\epsilon \otimes 1)%
\mathcal{F}=(1\otimes \epsilon )\mathcal{F}=1\,$where $\mathcal{F}_{12}=%
\mathcal{F}\otimes 1$ and $\mathcal{F}_{23}=1\otimes \mathcal{F}$.

The twist $\mathcal{F}$ does not modify the algebraic part and the counit,
but changes the coproducts $\Delta :\mathbb{H}\rightarrow \mathbb{H}\otimes 
\mathbb{H}$ and the antipodes $S:\mathbb{H}\rightarrow \mathbb{H}$ according
to formulae (\ref{ko1})-(\ref{ko2}). The quantum $\theta $-deformation is
generated by the twist (\ref{twistF}) . From the formula (\ref{ko1}) one
gets the coproducts 
\begin{align}
\Delta _{\mathcal{F}}(p_{\mu })& =p_{\mu }\otimes 1+1\otimes p_{\mu }
\label{Delta-p} \\
\Delta _{\mathcal{F}}(M_{\mu \nu })& =M_{\mu \nu }\otimes 1+1\otimes M_{\mu
\nu }  \label{M-Delta} \\
& -\frac{1}{2}\theta ^{\alpha \beta }[(\eta _{\alpha \mu }p_{\nu }-\eta
_{\alpha \nu }p_{\mu })\otimes p_{\beta }+p_{\alpha }\otimes (\eta _{\beta
\mu }p_{\nu }-\eta _{\beta \nu }p_{\mu })].  \notag
\end{align}%
From (\ref{twistF}) and (\ref{ko2}) follows that in considered case of $%
\theta _{\mu \nu }$-deformations $U=1$ and the antipodes remain unchanged,
i.e. $S_{\mathcal{F}}(h)=S_{0}(h)=-{h.}$

\subsection{Algebra of generalized coordinates $\mathbb{\hat{X}}$ as twisted 
$\mathbb{H}$-module}

For twist-deformed case we can introduce the deformed coordinates algebra $%
\mathbb{\hat{X}\ni }\hat{X}_{A}\mathbb{=}\{\hat{x}_{\mu },\hat{\Lambda}_{\mu
\nu }\}$ with the multiplication given by the star product formula 
\begin{equation}
\hat{X}_{A}\cdot \hat{X}_{B}\simeq X_{A}\star _{\mathcal{F}}X_{B}=m[\mathcal{%
F}^{-1}\circ (X_{A}\otimes X_{B})]=(\mathcal{F}_{(1)}^{-1}\rhd X_{A})(%
\mathcal{F}_{(2)}^{-1}\rhd X_{B})  \label{star}
\end{equation}%
where 
\begin{equation}
h\vartriangleright X_{A}=[h,X_{A}],\;\qquad h=\{p_{\mu },M_{\mu \nu
}\},\;X_{A}=\{x_{\mu },\Lambda _{\mu \nu }\}
\end{equation}%
and in undeformed case we obtain 
\begin{eqnarray}
\lbrack p_{\mu },x_{\nu }] &=&i\eta _{\mu \nu },\qquad \lbrack M_{\mu \nu
},x_{\rho }]=i(\eta _{\rho \nu }x_{\mu }-\delta _{\rho \mu }x_{\nu })
\label{non1} \\
\lbrack p_{\mu },\Lambda _{\rho \sigma }] &=&0,\qquad \lbrack M_{\mu \nu
},\Lambda _{\rho \sigma }]=\eta _{\rho \nu }\Lambda _{\mu \sigma }-\eta
_{\rho \mu }\Lambda _{\nu \sigma }.
\end{eqnarray}%
If we choose $X_{A}=f(X),X_{B}=g(X)$ the formula (\ref{star}) can be also
written as follows%
\begin{equation}
f(X)\star _{\mathcal{F}}g(X^{\prime })=\widehat{f(X)}\vartriangleright
g(X^{\prime })
\end{equation}%
where $\widehat{f(X)}$ denotes the noncommutative star representation of $f(%
\hat{X})$ defined by the formula (see also (\ref{ncco}))%
\begin{equation}
f(\hat{X})\simeq \widehat{f(X)}=m[\mathcal{F}^{-1}(\vartriangleright \otimes
1)(f(X)\otimes 1)]  \label{sttar}
\end{equation}%
For the twist (\ref{twistF}) we get from (\ref{sttar}) the following
explicit formulas describing generalized coordinates $\hat{X}_{A}=\{\hat{x}%
_{\mu },\hat{\Lambda}_{\mu \nu }\}$ in terms of undeformed relativistic
quantum phase space variables $(x_{\mu },p_{\mu })$ and $\Lambda _{\mu \nu }$
\begin{equation}
\hat{x}^{\mu }=x^{\mu }+\frac{1}{2}\theta ^{\mu \alpha }p_{\alpha },\qquad 
\hat{\Lambda}_{\rho \sigma }=\Lambda _{\rho \sigma }  \label{repa}
\end{equation}%
i.e. Lorentz group parameters remain classical. Due to (\ref{non1}) and (\ref%
{repa}) one gets the expected algebraic relations 
\begin{eqnarray}
&&\left[ \hat{x}_{\mu },\hat{x}_{\nu }\right] =i\theta _{\mu \nu }
\label{x1} \\
&&[\hat{x}_{\mu },\hat{\Lambda}_{\rho \sigma }]=[\hat{\Lambda}_{\mu \nu },%
\hat{\Lambda}_{\rho \sigma }]=0\,.  \label{x2}
\end{eqnarray}%
Using the relation%
\begin{equation}
h\vartriangleright (\hat{X}_{A}\hat{X}_{B})=(h_{(1)}\vartriangleright \hat{X}%
_{A})(h_{(2)}\vartriangleright \hat{X}_{B})\   \label{kkk}
\end{equation}%
following \cite{chaichian}\ one can check easily that the commutators (\ref%
{x1})-(\ref{x2}) are covariant under the action (\ref{kkk}) of $\theta _{\mu
\nu }$-deformed Poincare-Hopf algebra.

\subsection{$\protect\theta _{\protect\mu \protect\nu }$-deformed quantum
phase space $\mathcal{H}_{\protect\theta }^{(10+10)}=(\hat{x}_{\protect\mu },%
\hat{\Lambda}_{\protect\mu \protect\nu };p_{\protect\mu },M_{\protect\mu 
\protect\nu })$}

\bigskip Using (\ref{repa}) one can check the following set of cross
commutators%
\begin{eqnarray}
\lbrack p_{\mu },\hat{x}_{\nu }] &=&i\eta _{\mu \nu }  \label{mn1} \\
\lbrack p_{\mu },\hat{\Lambda}_{\rho \sigma }] &=&0 \\
\lbrack M_{\mu \nu },\hat{\Lambda}_{\rho \sigma }] &=&-i(\eta _{\rho \mu }%
\hat{\Lambda}_{\nu \sigma }-\eta _{\rho \nu }\hat{\Lambda}_{\mu \sigma }) \\
\lbrack M_{\mu \nu },\hat{x}_{\rho }] &=&i\eta _{\rho \nu }(\hat{x}_{\mu }-%
\frac{1}{2}\theta _{\mu }^{\;\alpha }p_{\alpha })-i\eta _{\rho \mu }(\hat{x}%
_{\nu }-\frac{1}{2}\theta _{\nu }^{\;\alpha }p_{\alpha })  \label{mn4} \\
&&-\frac{i}{2}(\theta _{\rho \mu }p_{\nu }-\theta _{\rho \nu }p_{\mu }). 
\notag
\end{eqnarray}%
Together with commutators (\ref{x1})-(\ref{x2}) the set of relations (\ref%
{mn1})-(\ref{mn4}) satisfies the Jacobi identities and defines the algebra
of $\theta _{\mu \nu }$-deformed quantum phase space $\mathcal{H}_{\theta
}^{(10+10)}$.

\section{$\protect\theta _{\protect\mu \protect\nu }$-deformed quantum
Poincare matrix group $\mathbb{G}$ and corresponding Heisenberg double $%
\mathbb{G\rtimes H}$}

\subsection{$RTT$ quantization method and $\protect\theta _{\protect\mu 
\protect\nu }$-deformed quantum Poincare group algebra $\mathbb{G}$}

The universal $\mathcal{R}$-matrix ($(a\wedge b=a\otimes b-b\otimes a)$) 
\begin{equation}
\mathcal{R}=\mathcal{F}^{T}\,\mathcal{F}^{-1}=\exp [-i\theta ^{\mu \nu
}p_{\mu }\otimes p_{\nu }]\,\qquad \qquad (a\otimes b)^{T}=b\otimes a\,
\label{dlww2.4}
\end{equation}%
can be used for the description of 10-generator deformed $D=4$ Poincare
group. Using the $5\times 5$ - matrix realization of the Poincare generators 
\begin{equation}
(M_{\mu \nu })_{\ B}^{A}=\delta _{\ \mu }^{A}\eta _{\nu B}-\delta _{\ \nu
}^{A}\eta _{\mu B}\,\qquad \qquad (p_{\mu })_{\ B}^{A}=\delta _{\ \mu
}^{A}\delta _{\ B}^{4}\,  \label{rep}
\end{equation}%
we can show that in (\ref{dlww2.4}) only the linear term is non-vanishing,
i.e.%
\begin{equation}
\mathcal{R}=1\otimes 1-i\theta ^{\mu \nu }p_{\mu }\otimes p_{\nu }.\,
\label{rrr}
\end{equation}

To find the matrix quantum group, which provides the Hopf algebra dual to $%
\mathbb{H}$ in the matrix realization (\ref{rep}), we introduce the
following $5\times 5$ - matrices 
\begin{equation}
\mathcal{\hat{T}}_{AB}=\left( 
\begin{array}{cc}
\hat{\Lambda}_{\mu \nu } & \hat{\xi}_{\mu } \\ 
0 & 1%
\end{array}%
\right) \,  \label{dlww2.7}
\end{equation}%
where $\hat{\Lambda}_{\mu \nu }$ parametrizes the quantum Lorentz rotation
and $\hat{\xi}_{\mu }$ denotes quantum translations. In the framework of the 
$\mathcal{FRT}$ procedure, the algebraic relations defining such a quantum
group $\mathbb{G}$ are described by the following relation 
\begin{equation}
\mathcal{R}\hat{\mathcal{T}}_{1}\hat{\mathcal{T}}_{2}=\hat{\mathcal{T}}_{2}%
\hat{\mathcal{T}}_{1}\mathcal{R}\,  \label{dlww2.8}
\end{equation}%
while the composition law for the coproduct remains classical $\Delta (\hat{%
\mathcal{T}}_{AB})=\hat{\mathcal{T}}_{AC}\otimes \hat{\mathcal{T}}_{\ B}^{C}$
with $\hat{\mathcal{T}}_{1}=\hat{\mathcal{T}}\otimes 1$, $\hat{\mathcal{T}}%
_{2}=1\otimes \hat{\mathcal{T}}$ and quantum $\mathcal{R}$-matrix (\ref{rrr}%
) given in the representation (\ref{rep}).\newline
In terms of the basis $(\hat{\xi}_{\mu },\hat{\Lambda}_{\mu \nu })$ of $%
\mathbb{G}$ the algebraic relations (\ref{dlww2.8}), describing the quantum
group algebra, can be written as follows 
\begin{eqnarray}
&&\left[ \hat{\xi}_{\mu },\hat{\xi}_{\nu }\right] =i\theta ^{\rho \sigma
}(\eta _{\mu \rho }\eta _{\nu \sigma }-\hat{\Lambda}_{\mu \rho }\hat{\Lambda}%
_{\nu \sigma })\,:=i\Theta _{\mu \nu }(\hat{\Lambda}),  \label{dlww2.10a} \\
&&[\hat{\xi}_{\mu },\hat{\Lambda}_{\rho \sigma }]=0\,,\qquad \lbrack \hat{%
\Lambda}_{\mu \nu },\hat{\Lambda}_{\rho \sigma }]=0\,  \label{dlww2.10b}
\end{eqnarray}%
while the coproduct takes the well known classical form 
\begin{equation}
\Delta \,(\hat{\Lambda}_{\mu \nu })=\hat{\Lambda}_{\mu \rho }\otimes \hat{%
\Lambda}_{\ \nu }^{\rho }\,\qquad \qquad \Delta (\hat{\xi}_{\mu })=\hat{%
\Lambda}_{\mu \nu }\otimes \hat{\xi}^{\nu }+\hat{\xi}_{\mu }\otimes 1\,.
\label{dlww2.11}
\end{equation}%
One can check that coproducts (\ref{dlww2.11}) are homomorphic to the
algebra (\ref{dlww2.10a})-(\ref{dlww2.10b}) defining $\theta _{\mu \nu }$%
-deformed quantum Poincare group.

\subsection{Duality between quantum Hopf algebras $\mathbb{H}$ and $\mathbb{G%
}$ and Heisenberg double $\mathcal{H=}\mathbb{H\rtimes G}$}

Two Hopf algebras \ $\mathbb{H},\mathbb{G}$\ are said to be dual if there
exists a nondegenerate bilinear form $\langle \,,\,\rangle \,:\mathbb{H}%
\mathcal{\times \mathbb{G}\longrightarrow }\mathbb{C},(h,\hat{g}%
)\longrightarrow \langle h,\,\hat{g}\,\rangle \,$such that the duality
relations 
\begin{eqnarray}
\langle h,\,\hat{g}\hat{g}^{\prime }\,\rangle \, &=&\langle \Delta (h),\,%
\hat{g}\otimes \hat{g}^{\prime }\,\rangle \,  \label{du1} \\
\langle hh^{\prime },\,\hat{g}\,\rangle \, &=&\langle h\otimes h^{\prime
},\Delta (\hat{g})\,\rangle \,.  \label{du2}
\end{eqnarray}%
are satisfied. In our considerations the following pairing relations%
\begin{equation}
<p_{\mu },\hat{\xi}_{\nu }>\ =i\eta _{\mu \nu }\quad <M_{\mu \nu },\hat{%
\Lambda}_{\alpha \beta }>\ =\ -i(\eta _{\mu \alpha }\eta _{\nu \beta }-\
\eta _{\nu \alpha }\eta _{\mu \beta })\quad <1,\hat{\Lambda}_{\mu \nu }>\ =\
\eta _{\mu \nu }  \label{pairing}
\end{equation}%
are appropriate. The basic action of $\mathbb{H}$ on $\mathbb{G}$ promoting $%
\mathbb{G}$ to the $\mathbb{H}$-module is given by the following relation%
\begin{equation}
h\blacktriangleright \hat{g}=\hat{g}_{(1)}\langle h,\hat{g}_{(2)}\rangle \ .
\end{equation}%
After using (\ref{du1}) one gets the relation%
\begin{eqnarray}
h\blacktriangleright (\hat{g}\hat{g}^{\prime }) &=&\hat{g}\hat{g}%
_{(1)}^{\prime }\langle \Delta h,\hat{g}_{(2)}\otimes \hat{g}_{(2)}^{\prime
}\rangle =\hat{g}\hat{g}_{(1)}^{\prime }\langle h_{(1)},\hat{g}_{(2)}\rangle
\langle h_{(1)},\hat{g}_{(2)}^{\prime }\rangle  \label{covv} \\
&=&(h_{(1)}\blacktriangleright \hat{g})(h_{(2)}\blacktriangleright \hat{g}%
^{\prime })\   \notag
\end{eqnarray}%
what establishes that algebra $\mathbb{G}$ is indeed the $\mathbb{H}$-module.

\bigskip In Heisenberg double framework we can obtain cross commutators
between the algebra $\mathbb{H}$ and group $\mathbb{G}$ by the following
relation%
\begin{equation}
\lbrack h,\hat{g}]=\hat{g}_{(2)}\langle h_{(1)},\hat{g}_{(1)}\,\rangle
\,h_{(2)}-\hat{g}h\qquad h=\{p_{\mu },M_{\mu \nu }\};\,\hat{g}=\{\hat{\xi}%
_{\mu },\hat{\Lambda}_{\mu \nu }\}.\qquad  \label{HD}
\end{equation}%
In such a way we obtain the quantum phase space algebra\footnote{%
For the case of $\kappa $-deformed quantum phase space see \cite{nowluk}.}.
Using pairing (\ref{pairing}), coproducts (\ref{Delta-p}), (\ref{M-Delta})
and formula (\ref{dlww2.11}) we get%
\begin{eqnarray}
\lbrack p_{\mu },\hat{\xi}_{\nu }] &=&i\eta _{\mu \nu }  \label{defo_1} \\
\lbrack p_{\mu },\hat{\Lambda}_{\rho \sigma }] &=&0  \label{defo2} \\
\lbrack M_{\mu \nu },\hat{\Lambda}_{\rho \sigma }] &=&-i(\eta _{\rho \mu }%
\hat{\Lambda}_{\nu \sigma }-\eta _{\rho \nu }\hat{\Lambda}_{\mu \sigma })
\label{defo_3} \\
\lbrack M_{\mu \nu },\hat{\xi}_{\rho }] &=&i\eta _{\rho \nu }(\hat{\xi}_{\mu
}-\frac{1}{2}\theta _{\mu }^{\;\alpha }p_{\alpha })-i\eta _{\rho \mu }(\hat{%
\xi}_{\nu }-\frac{1}{2}\theta _{\nu }^{\;\alpha }p_{\alpha })  \label{defo_4}
\\
&&-\frac{i}{2}(\theta _{\rho \mu }p_{\nu }-\theta _{\rho \nu }p_{\mu }) 
\notag
\end{eqnarray}%
The Hopf algebroid $\mathcal{\tilde{H}}^{(10+10)}=(\hat{\xi}_{\mu },\hat{%
\Lambda}_{\mu \nu };p_{\mu },M_{\mu \nu })$\ introduces an alternative model
of $\theta _{\mu \nu }$-deformed quantum phase space. It describes the
quantum phase space characterizing the dynamical system with the coordinates 
$(\hat{\xi}_{\mu },\hat{\Lambda}_{\mu \nu })$ which are specified by the NC
quantum Poincare group manifold $\mathbb{G}$.

\bigskip

\section{The covariance of $\mathbb{\hat{X}}$ under quantum Poincare group $%
\mathbb{G}$ and the generalized star product}

\subsection{\protect\bigskip The covariance of $\widehat{\mathbb{X}}$ under
quantum Poincare group $\mathbb{G}$}

We recall that $\widehat{\mathbb{X}}$ is the algebra of generalized
coordinates $\hat{X}_{A}=\{\hat{x}_{\mu },\hat{\Lambda}_{\mu \nu }\}$ and $%
\mathbb{G}$ is the algebra of Poincare symmetry parameters $\hat{g}=\{\hat{%
\xi}_{\mu },\hat{\Lambda}_{\mu \nu }\}$. One performs the quantum Poincare
transformations of $\hat{X}_{A}$ in the following way%
\begin{eqnarray}
\hat{x}_{\mu } &\longrightarrow &\hat{x}_{\mu }^{\prime }=\hat{\Lambda}_{\mu
\nu }\hat{x}^{\nu }+\hat{\xi}_{\mu }  \label{a} \\
\hat{\Lambda}_{\mu \nu } &\longrightarrow &\hat{\Lambda}_{\mu \nu }^{\prime
}=\hat{\Lambda}_{\mu }^{\;\alpha }\hat{\Lambda}_{\nu }^{\;\beta }\hat{\Lambda%
}_{\alpha \beta }=\hat{\Lambda}_{\mu \nu }.  \label{b}
\end{eqnarray}%
The commutators of algebra $\widehat{\mathbb{X}}$ (see (\ref{x1})-(\ref{x2}%
)) are invariant under a such transformation provided that the quantum
Poincare symmetry parameters $\mathbb{G}\ni \hat{g}=\{\hat{\xi}_{\mu },\hat{%
\Lambda}_{\mu \nu }\}$ satisfy the relations defining the algebra $\mathbb{G}
$ (see (\ref{dlww2.10a})-(\ref{dlww2.10b})) and $[\hat{X}_{A},\hat{g}]=0$.
In particular we have

\begin{equation}
\lbrack \hat{x}_{\mu },\hat{x}_{v}]=i\theta _{\mu \nu }\overset{\hat{x}_{\mu
}\rightarrow \hat{x}_{\mu }^{\prime }=\hat{\Lambda}_{\mu \nu }\hat{x}^{\nu }+%
\hat{\xi}_{\mu }}{\longrightarrow }[\hat{x}_{\mu }^{\prime },\hat{x}_{\nu
}^{\prime }]=i\theta _{\mu \nu }.
\end{equation}%
We see that the generalized coordinates algebra $\widehat{\mathbb{X}}$ in
different $\mathbb{G}$-frames specified by (\ref{a})-(\ref{b}) transform
covariantly under the transformations of quantum Poincare group $\mathbb{G}$.

In order to describe effectively the quantum Poincare transformations (\ref%
{a}) of NC functions $f(\hat{x}_{\mu })$ it is convenient to introduce the
generalized star products (see also \cite{klm}) representing the algebra of
functions $F(\hat{x}_{\mu },\hat{\xi}_{\mu })$ depending as well on NC
translations $\hat{\xi}_{\mu }$.

\subsection{Star product on the product $\mathbf{X}\mathbb{\otimes }\mathbf{T%
}$ with noncommutative translations of coordinates}

Let us consider firstly the star product $\star _{\mathcal{F}}^{^{\prime }}$
describing the NC product of algebra of functions $F(\hat{\xi}_{\mu })$
which depend on NC translations $\hat{\xi}_{\mu }\in \mathbb{G}$ satisfying
the relation (\ref{lnc})%
\begin{equation}
F(\hat{\xi}_{\mu })\cdot G(\hat{\xi}_{\nu })\simeq F(\xi _{\mu })\star _{%
\mathcal{F}}^{^{\prime }}G(\xi _{\nu })=m\circ \exp \left[ \frac{i}{2}\Theta
^{\mu \nu }(\hat{\Lambda})\frac{\partial }{\partial \xi ^{\mu }}\otimes 
\frac{\partial }{\partial \xi ^{\nu }}\right] F(\xi )\otimes G(\xi ),
\label{12star}
\end{equation}%
where in formula (\ref{12star}) one treats the Lorentz group parameters $%
\Lambda \equiv $ $\Lambda _{\mu \nu }$ as the numerical ones. Discussing the
quantum Poincare transformations (\ref{poin}) in field theory we should deal
with NC algebra of functions on $\mathbf{X}$ $(x_{\mu }\in \mathbf{X})$ as
well as on $\mathbf{T}$ $(\xi _{\mu }\in \mathbf{T})$, and subsequently use
the composite star product $\tilde{\star}_{\mathcal{F}}=\star _{\mathcal{F}%
}\cdot \star _{\mathcal{F}}^{^{\prime }}$ (see also \cite{koss2})%
\begin{eqnarray}
&&F(\hat{x}_{\mu },\hat{\xi}_{\mu })\cdot G(\hat{x}_{\nu },\hat{\xi}_{\nu
})\simeq F(x_{\mu },\xi _{\mu })\tilde{\star}_{\mathcal{F}}G(x_{\nu },\xi
_{\nu })  \label{star2} \\
&=&m\circ \exp \left[ \frac{i}{2}\left( \theta ^{\mu \nu }\frac{\partial }{%
\partial x^{\mu }}\otimes \frac{\partial }{\partial x^{\mu }}+\Theta ^{\mu
\nu }(\hat{\Lambda})\frac{\partial }{\partial \xi ^{\mu }}\otimes \frac{%
\partial }{\partial \xi ^{\mu }}\right) \right] (F(x_{\mu },\xi _{\mu
})\otimes G(x_{\nu },\xi _{\nu })).  \notag
\end{eqnarray}%
In formula (\ref{star2}) both relations (\ref{kom}), (\ref{lnc}) are taken
into account and we can represent the NC quantum Poincare transformations
given by $\mathbb{G}$ by using such a star product language. In particular
one can show that using the relation $\Lambda _{\mu \alpha }\Lambda _{\nu
\beta }\theta ^{\alpha \beta }+\Theta _{\mu \nu }(\Lambda )=\theta _{\mu \nu
}$ (see (\ref{lnc})) we get%
\begin{eqnarray}
&&F(\Lambda _{\mu \alpha }x^{\alpha }+\xi _{\mu })\tilde{\star}_{\mathcal{F}%
}G(\Lambda _{\nu \beta }x^{\beta }+\xi _{\nu })  \notag \\
&&\qquad \qquad =F(x_{\mu }^{\prime })\star _{\mathcal{F}}G(x_{\nu }^{\prime
})|_{x_{\mu }\rightarrow x_{\mu }^{\prime }=\Lambda _{\mu \alpha }x^{\alpha
}+\xi _{\mu };x_{\nu }\rightarrow x_{\nu }^{\prime }=\Lambda _{\nu \alpha
}x^{\alpha }+\xi _{\nu }} \\
&&\qquad \qquad =m\circ \exp \left[ \frac{i}{2}\theta ^{\mu \nu }\frac{%
\partial }{\partial x^{\prime \mu }}\otimes \frac{\partial }{\partial
x^{\prime \mu }}\right] (F(x_{\mu }^{\prime })\otimes G(x_{\nu }^{\prime
}))\equiv F(x_{\mu }^{\prime })\star _{\mathcal{F}}G(x_{\nu }^{\prime }). 
\notag
\end{eqnarray}%
In such a way we expressed in star product language the NC Poincare group
transformations by classical Poincare group transformations. We see that the
NC structure of quantum Poincare group translations are encoded in the
replacement of the star product $\star _{\mathcal{F}}\rightarrow \tilde{\star%
}_{\mathcal{F}}$. It can be shown by explicit calculation that three star
products $\star _{\mathcal{F}},\star _{\mathcal{F}}^{^{\prime }}$ and $%
\tilde{\star}_{\mathcal{F}}$ are associative.

\section{Two $\protect\theta _{\protect\mu \protect\nu }$-deformed Hopf
algebroids}

\subsection{Briefly on Hopf algebroids}

\bigskip The Hopf algebroids, introduced in \cite{lu} (see also \cite{20}),
are described by bialgebroids with supplemented antipodes. It has been
argued (see e.g. \cite{jkm}-\cite{lmw2018}) that the Hopf algebroids are
well adjusted to the description of physically important quantum (canonical
and noncanonical) phase space.

The bialgebroid $\mathcal{B}$ is specified by the set of the data $%
(H,A;s,t;m,\tilde{\Delta},\epsilon )$ where $H$ is the total algebra with
product $m$ and its subalgebra $A\subset H$ is called the base algebra. The
source map $s(a):A\rightarrow H$ is a homomorphism and the target map $%
t(a):A\rightarrow H$ \ an antihomomorphism, with their images commuting 
\begin{equation}
\lbrack s(a),t(b)]=0\qquad a,b\in A\qquad s(a),t(b)\in H.
\end{equation}%
The canonical choice of the source map is $s(a)=a$. One can introduce
natural $(A,A)-$bimodule structure on $H$ using source and target maps in
the basic $(A,A)$-bimodule formula, namely $ahb=ht(a)s(b)$. The coproducts $%
\tilde{\Delta}$ are described by the maps $H\rightarrow H\otimes _{A}H$ from 
$H$\ into $(A,A)$ bimodules $H\otimes _{A}H$, satisfying the
co-associativity condition%
\begin{equation}
(\tilde{\Delta}\otimes _{A}id_{H})\tilde{\Delta}=(id_{H}\otimes _{A}\tilde{%
\Delta})\tilde{\Delta}.
\end{equation}%
Because $H\otimes _{A}H$ as the codomain of coproducts $\tilde{\Delta}$ does
not inherit the algebra structure from $H\otimes H$, in order to have well
defined multiplication one introduces the submodule $H\times _{A}H\subset
H\otimes H$ defined by the Takeuchi coproduct \cite{tak}.

The algebra $H$ with the product $m$ and the coalgebra with Takeuchi
coproduct $\tilde{\Delta}$ are compatible, i.e.%
\begin{equation}
\tilde{\Delta}(hh^{\prime })=\tilde{\Delta}(h)\tilde{\Delta}(h^{\prime }).
\end{equation}%
The coproduct $\tilde{\Delta}$ in $H\times _{A}H$ can be expressed in terms
of standard tensor product $H\otimes H$ by the equivalence classes
satisfying the condition%
\begin{equation}
\tilde{\Delta}(h)\mathcal{I}_{L}(a)=0,\qquad \mathcal{I}_{L}(a)=t(a)\otimes
1-1\otimes s(a),  \label{iii}
\end{equation}%
where $\mathcal{I}_{L}$ define the left ideal in $H\otimes H.$ The
equivalence classes defined by (\ref{iii}) are parametrized by so-called
coproduct gauge (see e.g. \cite{lws}).

The counit map $\epsilon :H\rightarrow A$ is satisfying $\epsilon (1_{H})=$ $%
1_{A}$, and%
\begin{equation}
(\epsilon \otimes _{A}id_{H})\tilde{\Delta}=(id_{H}\otimes _{A}\epsilon )%
\tilde{\Delta}=id_{H}.
\end{equation}%
We get Hopf algebroids if we are able to introduce an antipode (bijective
map) $S:H\rightarrow H$ which is an algebra antihomomorphism and satisfies
the following properties 
\begin{eqnarray}
S(t) &=&s  \label{an1} \\
m[(1\otimes S)\circ \gamma \tilde{\Delta}] &=&s\epsilon =\epsilon
\label{an2} \\
m[(S\otimes 1)\circ \tilde{\Delta}] &=&t\epsilon S,  \label{an3}
\end{eqnarray}%
where in general case we need additional linear map $\gamma :H\otimes
_{A}H\rightarrow H\otimes H$, so-called anchor map.

\subsection{First choice: the coordinates $\widehat{\mathbb{X}}$ as the base
algebra (following \protect\cite{xu})}

Let us choose the bialgebroid group coproducts for base algebra generators $%
\hat{X}_{A}=\{\hat{x}_{\mu },\hat{\Lambda}_{\mu \nu }\}$%
\begin{equation}
\tilde{\Delta}(\hat{X}_{A})=\hat{X}_{A}\otimes 1  \label{kopp1}
\end{equation}%
satisfies the group and generalized quantum phase space algebra (\ref{mn1})-(%
\ref{mn4}).

The source $s(\hat{X}_{A})$ and the target $t(\hat{X}_{A})$ maps are the
following%
\begin{eqnarray}
s(\hat{X}_{A}) &=&m[\mathcal{F}^{-1}(\vartriangleright \otimes
1)(s_{0}(X_{A})\otimes 1)]=\hat{X}_{A}  \label{st1} \\
t(\hat{x}_{\mu }) &=&m[(\mathcal{F}^{-1})^{\tau }(\vartriangleright \otimes
1)(t_{0}(x_{\mu })\otimes 1)] \\
&=&x_{\mu }-\frac{1}{2}\theta _{\mu }^{\;\alpha }p_{\alpha }=\hat{x}_{\mu
}-\theta _{\mu }^{\;\alpha }p_{\alpha }  \notag \\
t(\hat{\Lambda}_{\mu \nu }) &=&m[(\mathcal{F}^{-1})^{\tau
}(\vartriangleright \otimes 1)(t_{0}(\Lambda _{\mu \nu })\otimes 1)]=\Lambda
_{\mu \nu }=\hat{\Lambda}_{\mu \nu }  \label{st3}
\end{eqnarray}%
where $s_{0}(X_{A})=X_{A}$, $t_{0}(X_{A})=X_{A}$. The maps (\ref{st1})-(\ref%
{st3}) satisfy the following relations 
\begin{eqnarray}
\lbrack s(\hat{X}_{A}),t(\hat{X}_{B})] &=&0 \\
\lbrack s(\hat{x}_{\mu }),s(\hat{x}_{\nu })] &=&i\theta _{\mu \nu },\qquad
\lbrack t(\hat{x}_{\mu }),t(\hat{x}_{\nu })]=-i\theta _{\mu \nu } \\
\lbrack s(\cdot ),s(\cdot )] &=&[t(\cdot ),t(\cdot )]=0\;\;(\text{for the
other choices of }\hat{X}_{A}).
\end{eqnarray}%
The counit has canonical form%
\begin{equation}
\epsilon (\widehat{X}_{A})=m[\mathcal{F}^{-1}(\triangleright \otimes
1)(\epsilon _{0}(X_{A})\otimes 1)]=\widehat{X}_{A}.
\end{equation}%
Using (\ref{an1}) one gets explicit formulae for the antipodes%
\begin{equation}
S(\widehat{X}_{A})=t(\widehat{X}_{A}).
\end{equation}%
In our case we have $S^{2}=1.$ One can check that for ideal $\mathcal{I}%
=t\otimes 1-1\otimes s$ it is true that%
\begin{equation}
m[(1\otimes S)]\mathcal{I}=m[(S\otimes 1)]\mathcal{I=}0
\end{equation}%
and it follows that we do not need the anchor map (see (\ref{an2})).

In order to determine the ideal (\ref{iii}) let us start form the
nondeformed ideal (for $\theta _{\mu \nu }=0$)%
\begin{equation}
\mathcal{I}_{0}(X_{A})=X_{A}\otimes 1-1\otimes X_{A}.
\end{equation}%
We can obtain $\widehat{X}_{A}$ from $X_{A}$\ by using twisting formula (\ref%
{sttar}). One gets for our twisted algebroid the following twist-deformed
ideal (see (\ref{st1})-(\ref{st3}))

\begin{equation}
\mathcal{I}_{\mathcal{L}}(\widehat{X}_{A})=\mathcal{FI}_{0}(X_{A})\mathcal{F}%
^{-1}=t(\widehat{X}_{A})\otimes 1-1\otimes s(\widehat{X}_{A}).
\end{equation}%
In particular one can check the following relations

\begin{equation}
\lbrack \tilde{\Delta}(\hat{X}_{A}),\mathcal{I}_{\mathcal{L}}(\widehat{X}%
_{A})]=[\mathcal{I}_{\mathcal{L}}(\widehat{X}_{A}),\mathcal{I}_{\mathcal{L}}(%
\widehat{X}_{B})]=0.  \label{comm7}
\end{equation}%
If we change the bialgebroid coproduct (\ref{kopp1}) by introducing the
following coproduct gauge transformation (see also \cite{lws1},\cite{jkm},%
\cite{bp})%
\begin{equation}
\tilde{\Delta}(\widehat{X}_{A})\rightarrow \tilde{\Delta}_{\lambda }(%
\widehat{X}_{A})=\tilde{\Delta}(\widehat{X}_{A})+\lambda \mathcal{I}_{%
\mathcal{L}}(\widehat{X}_{A})  \label{il22}
\end{equation}%
it follows from (\ref{comm7}) that the new bialgebroid coproduct (\ref{il22}%
) describes the homomorphism of the commutation relations (\ref{x1})-(\ref%
{x2}) extended by the phase space commutators (\ref{mn1})-(\ref{mn4}).

The formula (\ref{il22}) after substituting the formulae (\ref{st1})-(\ref%
{st3}) for source as well as target map and using the exponential
parametrization%
\begin{equation}
\hat{\Lambda}_{\mu \nu }\equiv (e^{\hat{\omega}})_{\mu \nu }=\eta _{\mu \nu
}+\hat{\omega}_{\mu \nu }+\frac{1}{2}\hat{\omega}_{\mu }^{\;\rho }\hat{\omega%
}_{\rho \nu }+\mathcal{O}(\hat{\omega}^{3}),
\end{equation}%
takes for example for $\lambda =-\frac{1}{2}$ the following explicit form%
\begin{eqnarray}
\tilde{\Delta}_{-\frac{1}{2}}(\hat{x}_{\mu }) &=&\frac{1}{2}(\hat{x}_{\mu
}\otimes 1+1\otimes \hat{x}_{\mu })-\frac{1}{2}\theta _{\mu }^{\;\nu }p_{\nu
}\otimes 1,  \label{ff1} \\
\tilde{\Delta}_{-\frac{1}{2}}(\hat{\Lambda}_{\mu \nu }) &=&\frac{1}{2}(\hat{%
\Lambda}_{\mu \nu }\otimes 1+1\otimes \hat{\Lambda}_{\mu \nu })  \label{ff2}
\\
&\equiv &\eta _{\mu \nu }1\otimes 1+\frac{1}{2}(\hat{\omega}_{\mu \nu
}\otimes 1+1\otimes \hat{\omega}_{\mu \nu })+\mathcal{O}(\hat{\omega}^{2}), 
\notag
\end{eqnarray}%
where (\ref{ff1}) describes the $\theta _{\mu \nu }$-deformation of the
symmetric formula $\tilde{\Delta}_{-\frac{1}{2}}(x_{\mu })=\frac{1}{2}%
(x_{\mu }\otimes 1+1\otimes x_{\mu })$\footnote{%
For a pair of free non-relativistic two-particle with the same masses the
coproduct $\tilde{\Delta}_{-\frac{1}{2}}(x_{\mu })$ describes the global
coordinates describing the center of mass (see \cite{lws1}, Sect. 4).}$.$

\subsection{Second choice: quantum group $\mathbb{G}$ as the base algebra
(following \protect\cite{lu})}

The half-primitive bialgebroid coproducts for $\hat{g}=\{\hat{\xi}_{\mu },%
\hat{\Lambda}_{\mu \nu }\}$%
\begin{equation}
\tilde{\Delta}(\hat{g})=\hat{g}\otimes 1
\end{equation}%
together with the coproducts (\ref{Delta-p})-(\ref{M-Delta}) satisfy the
Heisenberg double commutators (see (\ref{dlww2.10a})-(\ref{dlww2.10b}) and (%
\ref{defo_1})-(\ref{defo_4})).

The source $s(\hat{g})$ and the target $t(\hat{g})$ maps should be
consistent with base algebra in the following sense%
\begin{eqnarray}
\lbrack s(\hat{g}),t(\hat{g}^{\prime })] &=&0 \\
\lbrack s(\hat{\xi}_{\mu }),s(\hat{\xi}_{\nu })] &=&i\Theta _{\mu \nu }(s(%
\hat{\Lambda})),\qquad \lbrack t(\hat{\xi}_{\mu }),t(\hat{\xi}_{\nu
})]=-i\Theta _{\mu \nu }(t(\hat{\Lambda}))  \label{phas2} \\
\lbrack s(\cdot ),s(\cdot )] &=&[t(\cdot ),t(\cdot )]=0\qquad (\text{for the
other choices of }\hat{g}).
\end{eqnarray}%
where due to (\ref{dlww2.10a}) the relations (\ref{phas2}) describe
quadratic algebras. Subsequently, it can be easily shown that analogously to
(\ref{st1})-(\ref{st3}) one gets 
\begin{eqnarray}
s(\hat{g}) &=&\hat{g}  \label{s2s} \\
t(\hat{\Lambda}_{\mu \nu }) &=&\hat{\Lambda}_{\mu \nu },\qquad t(\hat{\xi}%
_{\mu })=\hat{\xi}_{\mu }-\Theta _{\mu }^{\;\alpha }(\hat{\Lambda})p_{\alpha
}.  \label{t2t}
\end{eqnarray}

The counit is 
\begin{equation}
\epsilon (\hat{g})=\hat{g}.
\end{equation}%
and the antipodes which are given by%
\begin{equation}
S(\hat{g})=t(\hat{g})
\end{equation}%
satisfy the required relations (\ref{an1})-(\ref{an3}). Similarly as in Sect
5.2 $S^{2}=1$ and we do not need the anchor map.

If we consider the ideal 
\begin{equation}
\mathcal{I}_{\mathcal{L}}(\hat{g})=t(\hat{g})\otimes 1-1\otimes s(\hat{g})
\end{equation}%
and use the formulae (\ref{s2s})-(\ref{t2t}) one can introduce as well the
counterpart of the coproduct gauge transformations (\ref{il22}) and (\ref%
{ff1})-(\ref{ff2}).

\section{ Outlook}

In this paper we considered the most popular in the literature Moyal quantum
deformation of space-time coordinates and the corresponding quantum-deformed
noncanonical phase-spaces, described by $\theta _{\mu \nu }$-deformation of
relativistic Heisenberg algebra. Our aim was to present the pair of $\theta
_{\mu \nu }$-deformed phase spaces in the language of Hopf algebroid, with
the extension of translational sectors $(\hat{x}_{\mu },p_{\mu })$ or $(\hat{%
\xi}_{\mu },p_{\mu })$ by the rotational Lorentz phase space coordinates $(%
\hat{\Lambda}_{\mu \nu },M_{\mu \nu })$, describing in relativistic particle
models the spin degrees of freedom.

There were introduced two different relativistic quantum phase spaces with
Hopf-algebroid structure \ described in two ways:

\begin{itemize}
\item[-] by twist deformation of classical canonical Heisenberg bialgebroid,
describing $\theta _{\mu \nu }$-deformed relativistic quantum phase space
with NC Minkowski space-time coordinates and dual commuting fourmomenta (see
e.g. \cite{20}-\cite{bp},\cite{xu}),

\item[-] by considering dual pairs of quantum-deformed Poincare-Hopf
algebras (see \cite{lws}) which define the Poincare-Heisenberg double as
semidirect (smash) product of quantum $\theta _{\mu \nu }$-deformed Poincare
group describing generalized coordinate sector and dual quantum Poincare
algebra which provides the generalized momenta sector (see e.g. \cite{lws},%
\cite{lmmw}).
\end{itemize}

These methods were applied in order to obtain two versions of $\theta _{\mu
\nu }$-deformed quantum phase spaces. We considered the $(10+10)$%
-dimensional generalized phase spaces, with generalized coordinates
described respectively by $(\hat{x}_{\mu },\hat{\Lambda}_{\mu \nu })$ and $(%
\hat{\xi}_{\mu },\hat{\Lambda}_{\mu \nu })$ and different noncommutativity
for $\hat{x}_{\mu }$\ and $\hat{\xi}_{\mu }$: first one characterized by
constant $\theta _{\mu \nu }$ (see (\ref{kom})), and second one with $%
\Lambda $-dependent $\Theta _{\mu \nu }(\hat{\Lambda})$, with commutation
relations quadratic in $\hat{\Lambda}$ (see (\ref{dlww2.10a})).

The Hopf algebroids introduce new class of quantum spaces $H$ endowed with
bialgebroid structure, suitable for the description of quantum manifolds
with symplectic structure. The bialgebroidal coproducts $\tilde{\Delta}$ can
be introduced in the framework of standard tensor products $H\otimes H$ as
algebras defined modulo so-called coproduct gauges \cite{lws1},\cite{lws}.
The coproduct gauge freedom can parametrize various classes of dynamical
particle models, with different values of physical parameters, but with the
same symplectic algebraic structure in $H$, in $H\otimes H$ \ and higher
multiparticle sectors. A simple example has been provided in \cite{lws1}
where it was shown that the coproduct gauges in the model describing pairs
of free NR particles with masses $m_{1},m_{2}$ depend on the mass ratio $%
\frac{m_{1}}{m_{2}}$, which is a physical parameter. It should be pointed
out however that at present the role of coproduct gauges e.g. in phase space
description of physically important interacting relativistic particles still
is not well understood.

Second question which could be studied is the description and classification
of infinitesimal quantum deformations of Hopf algebroids. In particular it
would be interesting to introduce for Hopf algebroids the notion analogous
to classical $r$-matrices for Hopf algebras, as well as the bialgebroidal
counterpart of Yang-Baxter and pentagon equations. Answer to the last
problem is linked with previous problem, i.e. the understanding of the
physical content of the notion of coproduct gauges.

\section*{Acknowledgements}

J.L. and M.W. have been supported by Polish National Science Center, project
2017/27/B/ST2/01902.


\bigskip

\begin{thebibliography}{99}
\bibitem{e1} M.P. Bronstein, \textit{Quantum theory of week gravitational
fields}, \textit{JETP} \textbf{9} (1936) 140157

\bibitem{e2} H.S. Snyder, \textit{Quantized space-time, Phys. Rev.} \textbf{%
D71} (1947) 38

\bibitem{e3} S. Majid, \textit{Hopf algebras for physics at the Planck
scale, Class. Quant. Grav.} \textbf{5} (1988) 1587

\bibitem{e4} J. Lukierski, A. Nowicki, H. Ruegg, V. Tolstoy, \textit{%
q-deformation of Poincare algebra, Phys. Lett.} \textbf{B264} (1991) 331

\bibitem{majid} S. Majid, \textit{Foundations of Quantum \ Groups},
Cambridge University Press (1995)

\bibitem{e6} A. Connes, \textit{Non-commutative Geometry}, Acad. Press, 1944

\bibitem{e7} S. Doplicher, K. Fredenhagen, J.E. Roberts, \textit{The quantum
structure of space time at the Planck scale and quantum fields, Comm. Math.
Phys.} \textbf{172} (1995) 187

\bibitem{e8} E. J. Beggs, S. Majid, \textit{Quantum Riemannian Geometry}, in
press.

\bibitem{oeck} R. Oeckl, \textit{Untwisting noncommutative R}$^{d}$\textit{\
and the equivalence of quantum field theories, Nucl.Phys.} \textbf{B581}
(2000) 559, \href{https://arxiv.org/pdf/hep-th/0003018.pdf}{hep-th/0003018}

\bibitem{chaichian} M. Chaichian, P.P. Kulish, K. Nishijima, A. Tureanu, 
\textit{On a Lorentz-invariant interpretation of noncommutative space-time
and its implications on noncommutative QFT, } \textit{Phys.Lett.} \textbf{%
B604} (2004) 98, \href{https://arxiv.org/pdf/hep-th/0408069.pdf}{%
hep-th/0408069}

\bibitem{drin} V. G. Drinfeld, \textit{Proc. of XXth Math. Congress,
Berkeley (1985), publ. Berkeley Press, vol. }\textbf{1} (1986) 798

\bibitem{bloh} Ch. Blohmann, \textit{Covariant realization of quantum spaces
as star products by Drinfeld twists, J.Math.Phys}. \textbf{44} (2003) 4736, 
\href{https://arxiv.org/pdf/math/0209180.pdf}{math/0209180}

\bibitem{abp} P. Aschieri, A. Borowiec, A. Pachol, \textit{Observables and
dispersion relations in $\kappa $-Minkowski spacetime, JHEP} \textbf{1710}
(2017) 152, \href{https://arxiv.org/pdf/1703.08726.pdf}{arXiv.1703.08726}

\bibitem{koss} C. Gonera, P. Kosinski, P. Maslanka, S. Giller, \textit{%
Space-time symmetry of noncommutative field theory, Phys.Lett.} \textbf{B622}
(2005) 192, \href{https://arxiv.org/pdf/hep-th/0504132.pdf}{hep-th/0504132}; 
\textit{Global symmetries of noncommutative space-time}, \textit{Phys.Rev.} 
\textbf{D72} (2005) 067702, \href{https://arxiv.org/pdf/hep-th/0507054.pdf}{%
hep-th/0507054}

\bibitem{nowluk} J. Lukierski, A. Nowicki, \textit{Heisenberg double
description of }$\kappa $\textit{-Poincare algebra and }$\kappa $\textit{%
-deformed phase space}, \textit{Proc. XXI Int. Coll. Group. Theor. Methods
in Physics, ed. V.K. Dobrev, H.D. Doebner, Heron Press, Sofia} p. 186 (1997)
q-alg/9702003

\bibitem{sou} J. M. Souriau, \textit{Structure des systemes dynamiques}, 
\textit{ed. Dunod, Paris}, 1970

\bibitem{kos} B. Kostant, \textit{Quantization and representation theory}, 
\textit{Springer Lecture Notes in Math. }\textbf{170} (1970) 87

\bibitem{klm} P. Kosinski, J. Lukierski, P. Maslanka, \textit{Local field
theory on kappa Minkowski space, star products and noncommutative
translations, Czech. J. Phys.} \textbf{50} (2000) 1283, \href{https://arxiv.org/pdf/hep-th/0009120.pdf}%
{hep-th/0009120}

\bibitem{lws1} J. Lukierski, Z. Skoda, M. Woronowicz,\textit{\ }$\kappa $%
\textit{-deformed covariant quantum phase spaces as Hopf algebroids, Phys.
Lett.} \textbf{B750} (2015) 401, \href{https://arxiv.org/pdf/1507.02612.pdf}{%
arXiv.1507.02612}

\bibitem{lws} J. Lukierski, Z. Skoda, M. Woronowicz, \textit{On Hopf
algebroid structure of }$\kappa $\textit{-deformed Heisenberg algebra, Phys.
Atom. Nucl. }\textbf{80} (2017) 576, \href{https://arxiv.org/pdf/1601.01590.pdf}%
{arXiv.1601.01590}

\bibitem{koss2} C. Gonera, P. Kosinski, P. Maslanka, S. Giller, \textit{%
Global symmetries of noncommutative field theory, J.Phys.Conf.Ser.} \textbf{%
53} (2006) 865

\bibitem{lu} J-H Lu, \textit{Hopf algebroids and quantum grupoids, Int.
Journ. Math. }\textbf{7} (1996) \textbf{47}, \href{https://arxiv.org/pdf/q-alg/9505024.pdf}%
{q-alg/9505024}

\bibitem{20} T. Brzezinski, G. Militaru, \textit{Bialgebroids, }$\times _{A}$%
\textit{-bialgebras and duality, J. Algebra} \textbf{251} (2002) 279, \href{https://arxiv.org/pdf/math/0012164.pdf}%
{math.QA/0012164}

\bibitem{jkm} T.~Juri\'{c}, D.~Kova\v{c}evi\'{c}, S.~Meljanac, $\kappa $%
\textit{-deformed phase space, Hopf algebroid and twisting, SIGMA }\textbf{10%
} (2014) 106, \href{https://arxiv.org/pdf/1402.0397.pdf}{arXiv:1402.0397}

\bibitem{sss} S. Meljanac, Z. Skoda, M. Stoji\'{c}, \textit{Lie algebra type
noncommutative phase spaces as Hopf algebroid, Lett. Math. Phys.} \textbf{107%
} (2017) 475, \href{https://arxiv.org/pdf/1409.8188.pdf}{%
arXiv.1409.8188[math.QA]}

\bibitem{bp} A. Borowiec, A. Pacho\l , \textit{Twisted bialgebroids versus
bialgebroids from a Drinfeld twist}, \textit{J. Phys. }\textbf{A50} (2017)
055205, \href{https://arxiv.org/pdf/1603.09280.pdf}{arXiv.1603.09280}

\bibitem{lmmw} J. Lukierski, D. Meljanac, S. Meljanac, D. Pikutic, M.
Woronowicz, Lie\textit{-deformed quantum Minkowski spaces from twists:
Hopf-algebraic versus Hopf-algebroid approach, Phys. Lett.} \textbf{B777}
(2018) 1, \href{https://arxiv.org/pdf/1710.09772.pdf}{arXiv.1710.09772}

\bibitem{lmw2018} J. Lukierski, S. Meljanac, M. Woronowicz, \textit{Quantum
twist-deformed $D=4$ phase spaces with spin sector and Hopf algebroid
structures, Phys.Lett.} \textbf{B789} (2019) 82, \href{https://arxiv.org/pdf/1811.07365.pdf}%
{arXiv.1811.07365}

\bibitem{tak} M. Takeuchi, \textit{Groups of algebras over }$A\otimes \bar{A}
$, \textit{J. Math. Soc. Japan }\textbf{29} (1977) 459

\bibitem{xu} Ping Xu, \textit{Quantum grupoids, Commun. Math. Phys. }\textbf{%
216} (2001) 539, \href{https://arxiv.org/pdf/math/9905192.pdf}{math/9905192}
\end{thebibliography}
\end{document}